\documentclass[prd,superscriptaddress,twocolumn,preprint,10pt]{revtex4}
\usepackage{graphicx}

\def\MeV {\mathop{\hbox{MeV}}}

\def\Re {\mathop{\hbox{Re}}}

\def\Tr {\mathop{\hbox{Tr}}}
\def\DU  {\mathop{{\cal D}\hbox{U}}}
\def\Dpsi {\mathop{{\cal D}\bar{\psi}{\cal D}\psi}}
\def\dd  {\mbox{d}}
\newcommand\detn[1]{\mbox{det}_{#1}}
\newcommand\tdetn[1]{\widetilde{\mbox{det}}_{#1}}

\newcommand{\beq}{\begin{equation}}
\newcommand{\eeq}{\end{equation}}
\newcommand{\beqa}{\begin{eqnarray}}
\newcommand{\eeqa}{\end{eqnarray}}

\begin{document}

\title{Lattice QCD at finite density via a new canonical approach}

\author{Andrei Alexandru}
\affiliation{Department of Physics and Astronomy, University of Kentucky,
Lexington KY 40506, USA}
\author{Manfried Faber}
\affiliation{Atomic Institute of the Austrian Universities, Nuclear Physics
Division, A-1040 Vienna, Austria}
\author{Ivan Horv\'{a}th}
\affiliation{Department of Physics and Astronomy, University of Kentucky,
Lexington KY 40506, USA}
\author{Keh-Fei Liu}
\affiliation{Department of Physics and Astronomy, University of Kentucky,
Lexington KY 40506, USA}

\begin{abstract}

We carry out a finite density calculation based on a canonical approach which
is designed to address the overlap problem. Two degenerate flavor simulations
are performed using Wilson gauge action and Wilson fermions on $4^4$ lattices, 
at temperatures close to the critical temperature $T_c\approx 170\MeV$ and
large densities (5 to 20 times nuclear matter density). In this region, we
find that the algorithm works well. We compare our results with those from
other approaches. 
\end{abstract}

\preprint{UK/05-07}

\maketitle

\section{Introduction}

The phase structure of QCD at finite temperature and finite density is
relevant for a variety of phenomena: from subtle modifications of
cross-section in high energy collisions of nuclei, to exotic states of
nuclear matter in neutron stars. Due to asymptotic freedom we can use
perturbation theory to study the quark-gluon plasma at sufficiently
large temperatures. However, the regions of interest for heavy ion
collision experiments and astrophysics are essentially
non-perturbative.  Numerical studies of QCD are extremely helpful in
providing a quantitative understanding of the phase structure in these
regions.  

\begin{figure}[b]
\includegraphics[width=10cm]{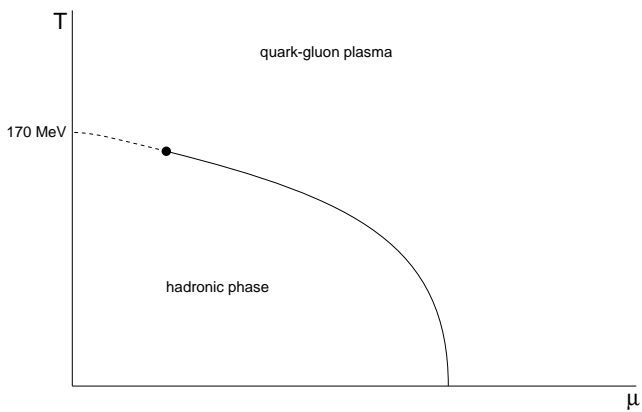}
\caption{Schematic picture of the expected QCD phase diagram.  The
solid line represents a first order phase transition, the dot a second
order phase transition and the dashed line represents a
crossover.\label{pict-phase-diagram}}
\end{figure}

At zero baryon density, it has been known for quite some time that QCD
undergoes a transition from a confined phase to a deconfined phase at
a temperature $T_c\approx 170\MeV$. Lattice QCD suggests that the
transition is in fact a smooth crossover. This is expected to turn
into a first order phase transition as the baryon density is
increased. A schematic picture of the expected phase diagram is
presented in Figure \ref{pict-phase-diagram}.

The position of the second order transition point, where the crossover
turns into a first order phase transition, is very important in providing
a quantitative understanding of the QCD phase diagram. 
Close to the critical temperature the relevant
degrees of freedom are the gluons and three flavors of quarks, the
light quarks, up and down, and the strange quark. The shape of the
curve seems to depend very little on the masses of the quarks, but the
position of the second order transition point depends strongly on the
mass of the strange quark. All numerical simulations treat the light
quarks as degenerate. If the strange quark mass is taken to be equal
to the mass of the light quarks, we have a theory with three
degenerate flavors. In this case, for low enough quark masses, the
zero density phase transition is expected to be first order, the
second order point disappears. As the strange quark mass is increased,
the zero density phase transition becomes a crossover, and the second
order phase transition point moves to larger and larger densities. As
the strange quark becomes infinitely heavy, only the light quarks
remain dynamically relevant.  The position of the second order phase
transition point is not the same as in the physically relevant case;
however, qualitatively the picture remains the same. This is why two
degenerate flavor QCD is interesting as a testbed for methods to
simulate finite density QCD.

Simulations at finite temperature and zero baryon density can be
performed using standard lattice techniques. However, non-zero baryon
density calculations remain one of the challenges of Lattice QCD.  The
reason is that, at non-zero chemical potential, the fermionic
determinant becomes complex and the standard Monte Carlo methods fail
since the integrand is no longer real and positive definite. The usual
approach is to split the integrand in two parts, one that is real and
positive and can be employed to generate an ensemble of configurations, and
another one that includes the complex phase of the determinant and is
folded into the observables. For clarity, let's write the grand canonical
partition function for Lattice QCD:
\beq
Z(V,T,\mu) = \int \DU \Dpsi e^{-S_g(U)-
S_f(\mu; U, \bar{\psi},\psi)},
\eeq
The fermionic part of the action, 
\beq
S_f(m, \mu; U, \bar{\psi},\psi) = \bar{\psi} M(m, \mu; U)\psi,
\eeq 
is a quark bilinear and we can perform the path integration analytically,
\beq
Z(V,T,\mu) = \int \DU e^{-S_g(U)} \prod_i \det M(m_i, \mu_i;U),
\eeq
where $M$ is the quark matrix and $m_i$ and $\mu_i$ are the mass and the 
chemical potential for flavor $i$. The gluonic part of the integrand,
$e^{-S_g(U)}$, is real and positive, whereas the fermionic part is only
guaranteed to be real when the chemical potential is zero. In the case of two 
degenerate flavors, after setting $\mu_1 = \mu_2 = \mu$, 
the partition function becomes
\beq
Z(V,T,\mu) = \int \DU e^{-S_g(U)} \det M(m, \mu; U)^2. \label{gc2flavours}
\eeq
The standard approach, the Glasgow reweighting method \cite{bar98},
 is to split the fermionic part into a real positive part
and a phase factor
\beq
\det M(\mu)^2 = \det M(\mu=0)^2\times \frac{\det M(\mu)^2}{\det M(\mu=0)^2},
\eeq
where we dropped some redundant indices. We can then apply the standard Monte
Carlo techniques to generate an ensemble according to the measure, 
\beq
P(U) = e^{-S_g(U)} \det M(\mu=0)^2,
\eeq
and then insert the phase factor,
\beq
\theta(\mu; U) = \frac{\det M(\mu; U)^2}{\det M(\mu=0; U)^2},
\eeq
into the observable
\beq
\left\langle O(U) \right\rangle_{\mu} = 
\frac{\left\langle O(U) \theta(\mu; U)\right\rangle_{\mu=0}}
{\left\langle \theta(\mu; U)\right\rangle_{\mu=0}}.
\eeq
To establish some terminology, we will call the ensemble generated with the
weight $P(U)$ the {\em generated ensemble} and, in a manner of speaking, we
will be calling {\em target ensemble} the one that would be generated
using the weight derived from the true action. For the second term, the word
ensemble is used loosely since for a complex integrand the concept of 
ensemble is, at best, ambiguous.

There are two major problems with the reweighting approach: the sign
problem and the overlap problem. The sign problem appears when the
phase factor, $\theta(\mu; U)$, averages to a value too close to zero
on the generated ensemble. By close, we mean an average value that is
smaller than the error. In that case, all the measurements will have
sizeable error bars and the method fails since we need extremely large
ensembles to get reasonable error bars. 

The second problem appears when the generated ensemble and the target
ensemble overlap poorly; for example, they might be in different
phases. More precisely, take an observable (in our example, the order
parameter for the phase transition), if it happens that the histogram
of this observable in the generated ensemble overlaps very poorly with
the histogram in the target ensemble, then the value that we get via
reweighting will be wrong. This problem is more serious than the sign 
problem since there is no indication when the measurement fails; the error bars
can be deceptively small \cite{bar98,cro01}.

Recently, a lot of progress has been made in studying the phase diagram
at temperatures around $T_c$ and small chemical potential
\cite{fod02,fod04,all03,all05,for02}. 
The reason
is that, in this region, the sign problem is manageable and the overlap
problem is expected to be under control.  The methods employ a more or
less sophisticated form of reweighting \cite{fod02,fod04,all03,all05}
or some form of analytical continuation 
from imaginary chemical potential \cite{for02}.  
The main results are
the shape of the phase transition curve around $T_c$ and the location
of the second order phase transition point for quark masses close to
the physical masses. All these simulations seem to be free of the sign
problem, but it is not clear whether the overlap problem is indeed
under control. One way to make sure that the results are correct
is to either use methods that are proved to be free of overlap
problem, or methods that are different enough but produce the same
results. For small values of $\mu/T_c \ll  1$, different methods seem to
agree. However, there is only one result for the location of the
second order phase transition point, and it occurs at rather large
value of the chemical potential. It is thus important to ask whether
this result is reliable.

In light of the problems mentioned above, it is imperative that new
methods be developed to simulate QCD at finite density. All the
methods mentioned above are based on the grand canonical partition
function. Far fewer attempts have been made to simulate QCD using the
canonical partition function \cite{eng99, kra03, kra04}. In this
paper, we will present simulations based on a method that employs the
canonical partition function \cite{liu02,liu03,ale04}. The main idea
is that, to avoid the overlap problem, it is essential to generate an
ensamble that is based on the projected determinant, instead of
reweighting. Moreover, to reduce the determinant fluctuations, the
updating process is broken into two steps: an HMC proposal and an
accept/reject step based on determinant ratios.  These runs are
exploratory in nature, using very small lattices and rather large
quark masses. The main goal is to determine the feasibility of the
algorithm and explore the available phase space.

The paper is organized as follows: in section II we will introduce the
canonical ensemble for QCD, in section III we present the algorithm we 
employed, in section IV we discuss the performance of the algorithm
and in section V we present the physical results. We then conclude
by attempting a physical interpretation of our results in section VI.

\section{Canonical partition function }

The simplest way to show how to build the canonical ensemble in 
Lattice QCD is to start from the fugacity expansion,
\beq
Z(V,T,\mu) = \sum_{n} Z_C(V, T, n) e^{\mu n/T},
\label{fugacity}
\eeq
where $n$ is the net number of quarks (number of quarks minus the
number of anti-quarks) and $Z_C$ is the canonical partition function. 
We note here that on a finite lattice, the
maximum net number of quarks is limited by the Pauli exclusion
principle.  Using the fugacity expansion, it is easy to see that 
we can write the canonical partition function as a Fourier transform 
of the grand canonical partition function,
\beq
Z_C(V, T, n) = \frac{1}{2\pi} \int_0^{2\pi} \mbox{d}\phi \,e^{-i n \phi} 
Z(V, T, \mu)|_{\mu=i\phi T}.
\eeq

We will now specialize to the case of two degenerate flavors. We use 
the grand canonical partition function in Eq. (\ref{gc2flavours}) to get
\beq
Z_C(V, T, n) = \int \DU e^{-S_g(U)} \detn{n} M^2(U), \label{zc}
\eeq
where we define
\beq
\detn{n} M^2(U) \equiv \frac{1}{2\pi}\int_0^{2\pi} \dd\phi\,e^{-i n \phi} 
\det M(m, \mu;U)^2|_{\mu=i\phi T}. \label{detproj}
\eeq
It is worth pointing out that the canonical partition function defined above
sums over configurations where the total net number of quarks, $n = n_1 + n_2$,
 is fixed. If we want to fix the net quark number for each flavor
then we would use $\detn{n_1} M(U)\times \detn{n_2} M(U)$.

For our study, we will be using Wilson fermions. To introduce a non-zero 
chemical potential, the fermion matrix at zero chemical potential,
\beqa
[M(U)]_{x,y} &=& \delta_{x,y} - \kappa \sum_{\mu=1}^4 (1-\gamma_\mu)U_\mu(x)
\delta_{x+\hat{\mu},y} \nonumber \\
&-& \kappa\sum_{\mu=1}^4 (1+\gamma_\mu) U_\mu^{\dagger}(y)\delta_{x,y+\hat\mu},
\label{fmatrix}
\eeqa
is altered \cite{has83} by introducing a bias for time 
forward propagation in the hopping matrix. More specifically, the
hopping in the time direction is altered
\beqa
(1+\gamma_4)U_4^{\dagger}(y)&\rightarrow& (1+\gamma_4)U_4^{\dagger}(y)
e^{\mu a}, \nonumber \\
(1-\gamma_4)U_4(x)&\rightarrow& (1-\gamma_4)U_4(x)e^{-\mu a}. 
\eeqa
We can perform a change of variables \cite{eng99},
\beqa
\psi(\vec{x},x_4)&\rightarrow& \psi'(\vec{x},x_4)=
e^{-\mu a x_4}\psi(\vec{x},x_4) \nonumber \\
\bar{\psi}(\vec{x},x_4)&\rightarrow& \bar{\psi}'(\vec{x},x_4)=
e^{\mu a x_4}\bar{\psi}(\vec{x},x_4) 
\eeqa
to restore the original form of the hopping matrix except on the last
time slice.  In terms of these new variables, we can write the
fermionic matrix as
$M(U_\phi) = M(m, \mu; U)$ where $M$ is defined in Eq. (\ref{fmatrix}) and
\beq
(U_\phi)_\nu(x)\equiv\left\{ 
\begin{array}{ll}
U_\nu(x) e^{-i\phi} & x_4=N_t ,\, \nu=4 \\
U_\nu(x) & \mbox{otherwise} .
\end{array}
\right.
\eeq
This should not be viewed as a change of the gauge field variables but
rather as a convenient way to write the fermionic matrix.

In order to evaluate numerically the partition function in Eq. (\ref{zc}), we
need to replace the continuous Fourier transform in Eq. (\ref{detproj}) with a
discrete one. We will then redefine the projected determinant,
\beq
\tdetn{n} M^2(U) \equiv \frac{1}{N} \sum_{j=0}^{N-1} 
e^{-i n \phi_j} \det M(U_{\phi_j})^2 . \label{detprojdisc}
\eeq
where $\phi_j = \frac{2\pi j}{N}$ and 
the parameter $N$ defines the discrete Fourier transform. In the limit
$N\rightarrow\infty$, we recover the original projected determinant. For finite
$N$ the partition function
\beq
\tilde{Z}_C(V,T,n) \equiv \int \DU e^{-S_g(U)} \tdetn{n} M^2(U), \label{tzc}
\eeq 
will only be an approximation of the canonical 
partition function. Using the fugacity expansion we can show that
\beq
\tilde{Z}_C(V,T,n)=\sum_{m=-\infty}^\infty Z_C(V,T,n+m N).
\eeq
If $N$ and $n$ are chosen such that $|n+mN|$ is minimal for $m=0$ then 
$\tilde{Z}_C$ should be a good approximation to $Z_C$ as long as
\beq
\frac{Z_C(V,T,n+mN)}{Z_C(V,T,n)} \ll 1,
\eeq
for all $m\not=0$. To understand better this condition take $0\leq n < N/2$;
the largest contamination comes from $Z_C(V,T,N-n)$. The ratio above is
\beq
\frac{Z_C(V,T,N-n)}{Z_C(V,T,n)}=e^{-\frac{F(V,T,N-n)-F(V,T,n)}{T}},
\eeq
where $F$ is the free energy of the system. For low temperatures, we expect 
that $F(V,T,n)\propto e^{-M_B |n|/3}$, where $M_B$ is the mass of the baryon.
We see then that the approximation will hold as long as the temperature is
low enough or the baryon mass is high. This assumption needs to be checked 
in our simulations; if it fails then we need to increase $N$.

\section{Algorithm}

\begin{figure*}[t]
\includegraphics[width=18cm]{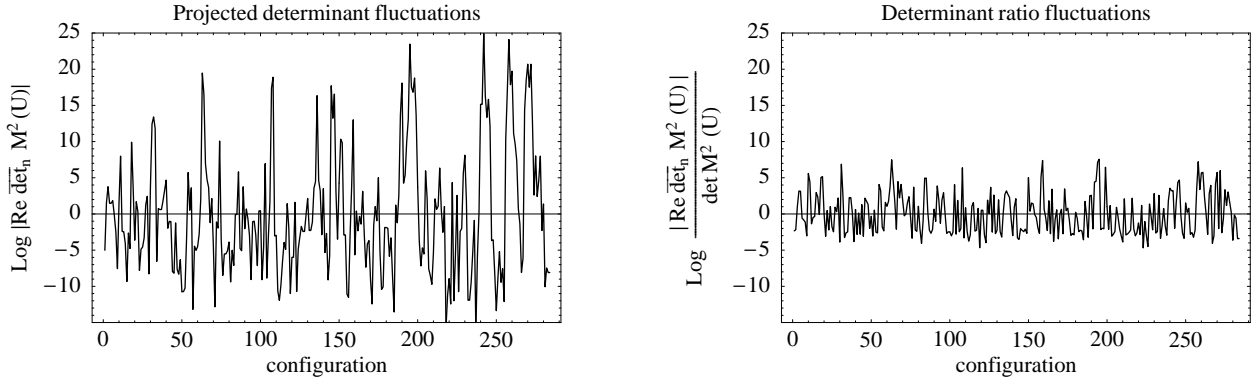}
\caption{Fluctuations of the fermionic part of the measure and of the
accept/reject factor $\omega(U)$, defined in Eq. (\ref{omega}),
as measured on an ensemble generated 
at $\beta=5.2$ and $n=3$. In both figures, we subtracted the average 
value so that the plots are centered around zero.
 \label{fig-fluctuations}}
\end{figure*}

In this section, we will present the algorithm we employ to simulate the
partition function $\tilde{Z}_C$.
Directly simulating the projected determinant in Eq. (\ref{detprojdisc}) is
known to face a fluctuation problem \cite{joo03}, since $\ln \det M=\Tr \ln M$
and $\Tr\ln M$ is proportional to the lattice volume. To alleviate the
problem we split the Markov process in two steps: a proposal step based on HMC
and an accept/reject step based on the ratio of the projected determinant
to the determinant used in the HMC step. Since the accept/reject is based on 
the determinant ratio, the fluctuations should be reduced and the acceptance
rate enhanced. We shall test this numerically.

\subsection{Target measure}

 To evaluate $\tilde{Z}_C$ using Monte
Carlo techniques, we need the integrand to be real and positive. Using 
$\gamma_5$ hermiticity of the fermionic matrix, i.e.
\beq
\gamma_5 M(U_\phi) \gamma_5 = M(U_\phi)^\dagger,
\eeq
we can easily prove that $\det M(U_\phi)$ is real. This implies that
\beq
\tdetn{n} M^2(U) = (\tdetn{-n} M^2(U))^*, \label{conj}
\eeq
but it doesn't imply that
the projected determinant is real. $\tdetn{n} M^2(U)$ is real only 
if $\det M(U_\phi) = 
\det M(U_{-\phi})$, which is not true configuration by configuration.
We can prove that for a charge conjugation symmetric action, this property
is true when averaged over the ensemble, i.e. 
$\left<\det M(U_\phi)\right> = \left<\det M(U_{-\phi})\right>$. In fact,
using charge conjugation symmetry of the action, we can prove that
\beq
\tilde{Z}_C(V,T,n) = \tilde{Z}_C(V,T,-n).
\eeq
This property allows us to rewrite the partition function
\beq
\tilde{Z}_C(V,T,n) = \int \DU e^{-S_g(U)} \Re \tdetn{n} M^2(U).
\eeq
Now the integrand is real but not necessarily positive. For the sake
of the argument, let's set aside for a while the fact that the integrand may
be negative and assume that the above expression can be evaluated using
standard Monte Carlo techniques. Even then, the fact
that we can write the partition function using a real integrand is not
sufficient. The goal of any simulation is to compute different observables,
the partition function itself is not of much interest. If we are only 
interested in observables that are even under charge conjugation, then
we could use the ensemble generated by the above action to compute
them. For observables that are odd under charge conjugation an additional
step is necessary: we have to reintroduce a phase. 

We want to emphasize that, if the above integrand is positive, the
observables which are even under charge conjugation could be evaluated directly
on the ensemble generated by the above action. Thus, we would have no
reweighting involved, and no overlap problem. The observables that are 
odd under charge conjugation are not guaranteed to behave as well, but
we assume that their behavior would be similar. At the worst, 
the extra phase might introduce a sign problem. 

We come back now to the positivity question. In the case that the integrand is 
not positive, we are forced to use the absolute value of the 
integrand as measure for our generated ensemble. The algorithm will then 
be designed to generate an ensemble according to the weight
\beq
W(U) = e^{-S_g(U)} \left|\Re \tdetn{n} M^2(U)\right|.
\eeq
The sign will be folded into the observables. For a generic observable the
sign will turn out to be some complex phase,
\beq
\alpha(U) = \frac{\tdetn{n} M^2(U)}{\left|\Re \tdetn{n} M^2(U)\right|},
\label{alpha}
\eeq
but for observables even under charge conjugation it will be just the
sign of $\Re \tdetn{n} M^2(U)$.

From the above discussion, it is clear that as long as we don't have a 
sign problem, the results of our simulation for observables invariant
under charge conjugation are reliable. For the other observables, the sign 
problem might be more severe, but we expect that they will not have an
overlap problem.

\subsection{HMC update}

Turning to the algorithmic issues, our approach to generating an
ensemble with weight $W(U)$ is to employ a Metropolis accept/reject
method. In short, the method employs a generating mechanism that
proposes new configurations with weight $W'(U)$ and then an
accept/reject step is used to correct for the target weight. Ideally,
the proposal mechanism would propose configurations with the weight
$W(U)$; in that case all new proposals will be accepted. In practice,
it isn't always possible to design efficient proposal mechanism for every
weight. The general approach is to use an efficient proposal mechanism
to generate a weight $W'(U)$ close to the target weight $W(U)$. If
successful, the acceptance rate would be high and the algorithm would
be efficient.

One possible solution is to use a heat-bath method to propose new
configurations based on the weight $W'(U)=e^{-S_g(U)}$. However, such
an updating strategy would be inefficient since the fermionic part is 
completely disregarded in the proposal step. The determinant, being an
extensive quantity, can fluctuate wildly from one configuration to the
next in the pure gauge updating process \cite{joo03,ale02}. 
To reduce the fluctuations, it was suggested \cite{liu03} that we should
employ an HMC algorithm for the proposal step. In this case
\beq
W'(U) = e^{-S_g(U)} \det M(U)^2.
\eeq
We will then accept the new configuration $U'$ with the probability
\beq
P_{acc} = \min\{1, \omega(U')/\omega(U)\},
\eeq
where $\omega$ is the ratio of the weights
\beq
\omega(U)=\frac{W(U)}{W'(U)} = \frac{\left|\Re\tdetn{n}M^2(U)\right|}
{\det M^2(U)}. \label{omega}
\eeq
We expect that this proposal mechanism will be more efficient.
Although the fermionic
part of the measure $\left|\Re\tdetn{n}M^2(U)\right|$
varies significantly from one configuration
to the next, the determinant ratio
\beq
\omega(U) = \frac{1}{N} \left| \sum_{j=0}^{N-1} \cos(\phi_j n)
e^{\Tr(\log M(U_{\phi_j}) - \log M(U))}\right|
\eeq
is expected to fluctuate less. We base our expectation on the fact
that, in the ratio, the leading fluctuations are removed by the $\Tr
\log$ difference of the quark matrices $M(U_{\phi_j})$ and $M(U)$.  To
check this, we compare the fluctuations in the fermionic part of the
determinant and the determinant ratios in one of the ensembles
generated in our runs.  This is shown in Fig. \ref{fig-fluctuations}.
We see that the fluctuations are significantly reduced which results
in a large boost in acceptance rate.

\subsection{Triality}

\begin{figure}[t]
\includegraphics[width=8cm]{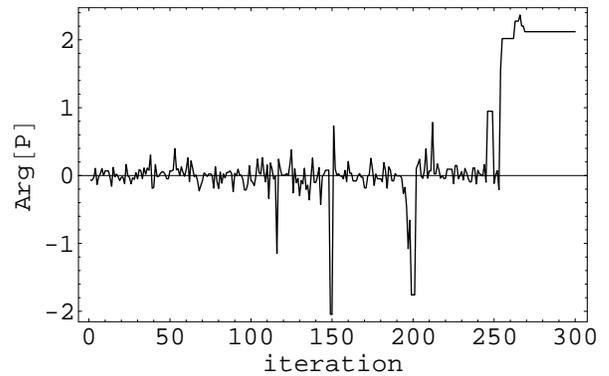}
\caption{Polyakov loop argument as a function of the simulation time.
Note that toward the end the value is unchanged for almost 50 iterations.
 \label{fig-z3problem}}
\end{figure}

The canonical partition function, Eq. (\ref{zc}), has a $Z_3$ symmetry
\cite{fab95} that is a direct consequence of the $Z_3$ symmetry of the
grand canonical partition function at imaginary chemical potential
\cite{rob86}. Under a transformation $U\rightarrow U_{\phi}$ with
$\phi=\pm 2\pi/3$, the gauge part of the action is invariant and 
\beq
\detn{n} M^2(U) \rightarrow \detn{n} M^2(U_{\pm 2\pi/3}) =
e^{\pm i \frac{2\pi}{3}n}\detn{n}M^2(U).
\eeq
We see then that when $n$ is a multiple of $3$ this transformation leaves
$\detn{n} M^2(U)$ invariant. Consequently, the canonical partition action
is invariant under this transformation. Incidentally, this symmetry of
the gauge part of the action together with the transformation above of
the fermionic part guarantees that the canonical partition function will
vanish when $n$ is not a multiple of $3$. However, this is no longer true if
this symmetry is spontaneously broken, which is  the case in the
deconfined phase. In this phase, there is no reason to expect that the 
canonical partition function should vanish when $n$ is not a multiple of $3$.

The transformation rule above is preserved for the discrete case if we
choose $N$, the parameter that defines the Fourier transform, to be a 
multiple of $3$. In our simulations, we will always choose $N$ to satisfy
this condition.
In this case, the remarks we made about $Z_C$ are valid 
for $\tilde{Z}_C$ and the projected determinant, $\tdetn{n} M^2(U)$ is 
invariant. Thus, the measure is symmetric under this transformation, i.e.
\beq
W(U) = W(U_{\pm 2\pi/3}).
\eeq
However, the HMC weight, $W'(U)$, does not have this symmetry since 
$\det M(U)$ is not invariant under this transformation. Because of this,
our algorithm can become frozen for long periods of time. For example,
in Fig. \ref{fig-z3problem}, we show how the argument of the Polyakov loop
changes with the simulation time if we use the method presented so far. 
We notice that at the end of the simulation, when
we tunnel to the sector where $\arg[P] \approx 2\pi/3$, the update is frozen;
the new proposals are rejected for a long time. This is due to the fact 
that HMC strongly prefers the $0$ sector. To understand this better, 
assume that we have a configuration $U_0$ in the $0$ sector, where 
$\arg[P(U_0)] \approx 0$, and denote with $U_+$ the configuration 
$(U_0)_{2\pi/3}$  with $\arg[P(U_+)] \approx 2\pi/3$. 
Then, we expect that $\det M^2(U_0)\gg\det M^2(U_+)$ since HMC prefers the $0$
sector, but $\tdetn{n} M^2(U_0) = \tdetn{n} M^2(U_+)$ since the projected
determinant is symmetric under the $Z_3$ transformations.  Assume now that 
HMC proposes $U_+$, the accept/reject step will accept this since
\beq
\frac{\omega(U_+)}{\omega(U_0)} = \frac{\det M^2(U_0)}{\det M^2(U_+)}\gg 1.
\eeq
However, in the next step HMC is likely to propose a new configuration in the
$0$ sector since it favors it strongly. By the reverse of the argument above
we have that
\beq
\frac{\omega(U_0)}{\omega(U_+)} = \frac{\det M^2(U_+)}{\det M^2(U_0)}\ll 1
\eeq
and the new configuration will be very likely rejected.

This means that although the algorithm will end up sampling the three
sectors equally, as required by the symmetric weight $W(U)$, two
of the sectors will take a very long time to sample properly. To
address this problem, we introduce a $Z_3$ hopping \cite{kra03}. Since
the weight $W(U)$ is symmetric under the $Z_3$ transformation, we can
intermix the regular updates with a change in the field variables
$U\rightarrow U_{\pm 2\pi/3}$. We will choose the sign randomly, with
equal probability for each sign, to satisfy detailed balance.  The new
algorithm will sample all sectors in the same manner.

\section{Simulation details}

Most of the computer time in these simulations is spent computing the
determinant. There is a proposal that would employ a determinant estimator
\cite{thr98}, but in this work we compute the determinant exactly using LU
decomposition. This is a very expensive calculation considering that even
for the small lattices we used in this study the fermionic matrix has $3072$
rows. Furthermore, the algorithm scales with the third power of the lattice
four volume and it is not easily parallelizable. The high 
computational cost constrains us to use only $4^4$ lattices for this study. 

The computational cost increases linearly with the parameter $N$ used to
define the Fourier transform. For our study, we used $N=12$. For each value
of $\beta$ we run three simulations: $n=0$, $n=3$, and $n=6$. They correspond
to $0$, $1$, and $2$ baryons in the box.

Since our volume in lattice units is small, we had to use large lattice
spacings.  We had runs for $\beta=5.00, 5.10, 5.15, 5.20, 5.25, 5.30,
\text{ and } 5.35$ and we fixed $\kappa=0.158$. The relevant
parameters can be found in Table~\ref{table-run-param}. The lattice 
spacing and the pion mass are determined using standard dynamical action
on a $12^4$ lattice for the same values of $\beta$ and $\kappa$. 
The lattice spacing was determined by using $r_0$ scale\cite{som94}. We note
that the pion mass varies very little with $\beta$, consequently the 
quark mass is roughly the same in all runs. We also note that the quark
mass is quite heavy, above the strange quark mass.

\begin{table}
\caption{Simulation parameters. \label{table-run-param}}
\begin{tabular}{|c|c|c|c|c|}
\hline
$\beta$ & a(fm) & $m_\pi$(MeV) & $V^{-1}(fm^{-3})$ & T(MeV) \\
\hline
5.00 & 0.343(2) & 926(7) & 0.387(7) & 144(1) \\
5.10 & 0.322(4) & 945(13) & 0.468(17) & 153(2) \\
5.15 & 0.313(3) & 942(11) & 0.510(15) & 157(2) \\
5.20 & 0.300(1) & 945(5)  & 0.579(6)  & 164(1) \\
5.25 & 0.284(5) & 945(20) & 0.682(36) & 173(3) \\
5.30 & 0.260(1) & 973(9)  & 0.889(10) & 189(1) \\
5.35 & 0.233(2) & 959(14) & 1.235(32) & 211(2) \\
\hline
\end{tabular}

\end{table}

For the HMC update, we used the $\Phi$ algorithm \cite {got87} made
exact by an accept/reject step at the end of each trajectory\cite{dua87}. For
updating process, we set the length of the trajectories to $0.5$ with
$\Delta\tau = 0.01$. The HMC acceptance rate was very close to $1$ since
the step length was very small. We adjust the number of HMC trajectories
between two consecutive finite density accept/reject steps so that the
acceptance rate stays in the range 15\% to 30\%. The relevant
information is collected in Table \ref{table-acceptance}. We see that
we can get decent acceptance rates even when consecutive finite density
Metropolis steps are quite far apart in configuration space. This allows us 
to move very fast through the configuration space. We collected about 
100 configurations for each run, separated by 10 accept/reject steps.

\begin{table}[b]
\caption{Acceptance rates; we list first the number of HMC trajectories
between two consecutive finite density Metropolis steps
and then the acceptance rate. \label{table-acceptance}}
\begin{tabular}{|c|cc|cc|cc|}
\hline
$\beta$ & HMC traj & n=0 & HMC traj & n=3 & HMC traj & n=6  \\
\hline
5.00 & 50 & 0.59(2) & 20 & 0.27(1) & 02 & 0.19(1)  \\
5.10 & 50 & 0.55(2) & 20 & 0.29(2) & 05 & 0.15(1)  \\
5.15 & 50 & 0.53(1) & 20 & 0.25(2) & 05 & 0.18(1)  \\
5.20 & 50 & 0.49(2) & 20 & 0.26(2) & 05 & 0.25(2)  \\
5.25 & 50 & 0.40(2) & 20 & 0.32(1) & 05 & 0.40(2)  \\
5.30 & 50 & 0.36(1) & 50 & 0.34(2) & 10 & 0.32(2)  \\
5.35 & 50 & 0.33(2) & 50 & 0.34(2) & 10 & 0.38(1)  \\
\hline
\end{tabular}

\end{table}

From an algorithmic point of view, one of the most interesting
questions is whether or not we have a sign problem. To settle this
question, we measured the average phase $\alpha(U)$ given in
Eq. (\ref{alpha}). In fact, it is easy to prove that the imaginary part
of the phase should vanish on the ensemble average. It is the real
part of this phase that carries the signal of a sign problem; if its
average is close to zero then we have a sign problem. We note here that
the real part of the phase is $\pm 1$, and that the sign problem appears
when we have an almost equal number of configurations of each sign.

\begin{figure}
\includegraphics[width=9cm]{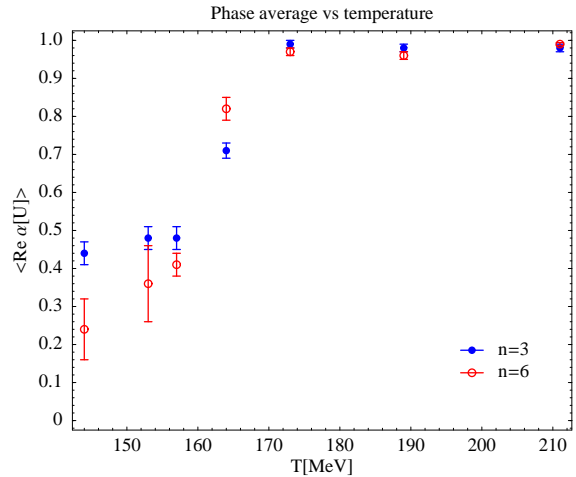}
\caption{Average phase for $n=3$ and $n=6$ runs. The $n=0$ runs have all signs
equal to $1$ so we did not plot them in this figure.
 \label{fig-sign-problem}}
\end{figure}

In Fig. \ref{fig-sign-problem}, we plot the average of the real part
of the phase $\alpha$ as a function of the temperature. We note that
in the deconfined phase, the projected determinant is positive most of
the time; as we go into the hadronic phase the sign starts
oscillating. Deep in the hadronic phase, the oscillations are more
severe at higher density which we can see by comparing the case of
$n=6$ with $n=3$ in Fig. \ref{fig-sign-problem}.  However, it is
possible that at $T<T_c$ the sign average is actually smaller at lower
density.  This is due to the fact that, at this temperature, it is
possible to have the system in the hadronic phase at low densities and
in the quark-gluon plasma phase at higher densities. Since the
oscillations are more severe in the hadronic phase it is not
surprising that close to and below $T_c$ we would have more sign
oscillations at lower density. This could explain the average sign
reversal of $n=3$ and $n=6$ at $T=164\MeV$ as compared to those at 
other temperatures in Fig. \ref{fig-sign-problem}.

In Fig. \ref{fig-sign-problem}, we also see that the sign average drops
sharply as we go through the transition temperature but the rate slows down,
as we go deeper in the hadronic phase. This slowing down may be due to the
fact that as we go to lower and lower temperatures, the physical
volume of the box is also increased and the density decreases. 

In conclusion, it seems that, at least for a $4^4$ lattice, we should
be able to investigate the region where $T>0.8T_c$ and baryon number
$n_B<3$.  We listed here the baryon number since we expect that the
sign fluctuations are going to be determined by this number rather
than the baryon density.  From Table \ref{table-run-param} we see that
the densities used in this study are rather large; they range from
$2.4$ to $24$ times the nuclear matter density.  An interesting future
direction would be to increase the spatial volume, using for example a
$6^3\times 4$ lattice, while keeping the baryon number the same.  This
would allow us to study densities closer to the physically interesting
region. If the sign oscillation is really determined by the baryon
number not volume, we should be able to use this algorithm to explore
this region. It is also clear that a sign problem will appear at
baryon numbers larger than the ones employed in this study.  We also show
that the algorithm can be efficient in going through the configuration
space. For non-zero density runs, the acceptance rate drops quite
significantly with the temperature; much smaller number of
trajectories have to be used between the successive accept/reject
steps. This may be due to a decrease in the autocorrelation time; as
we go to smaller values of $\beta$ the autocorrelation is expected to
decrease. A more detailed study is needed to quantify this statement.

\section{Physical results}

We turn now toward the physical results. We will present measurements
of the Polyakov loop, chemical potential, chiral condensate and the
conserved charge. We feel compelled to point out that the results
presented here have large systematic errors. The lattice volume
and the baryon number are small, consequently the finite size effects
are going to be important.  The lattice spacing is very large and the
lattice artifacts will be substantial.  Since we are using Wilson
fermions we expect that the chiral symmetry is broken quite badly by
lattice terms. Also, the quark mass is rather heavy. In light of these
problems, the results presented in this section are interesting more as
proof of concept results.

\begin{figure}[t]
\includegraphics[width=6cm,angle=-90]{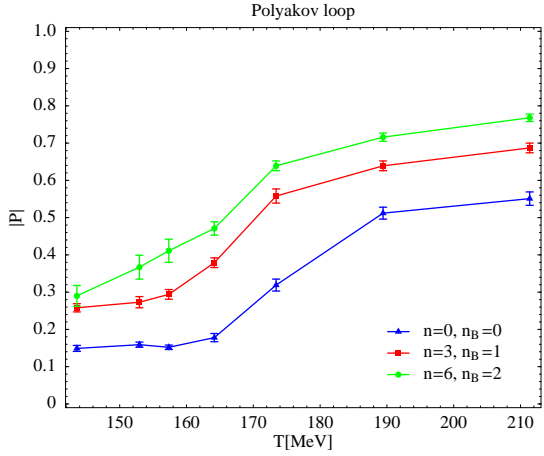}
\caption{Polyakov loop as a function of temperature. \label{fig-poly}}
\end{figure}

\subsection{Polyakov loop}

\begin{figure}[t]
\includegraphics[width=9.5cm]{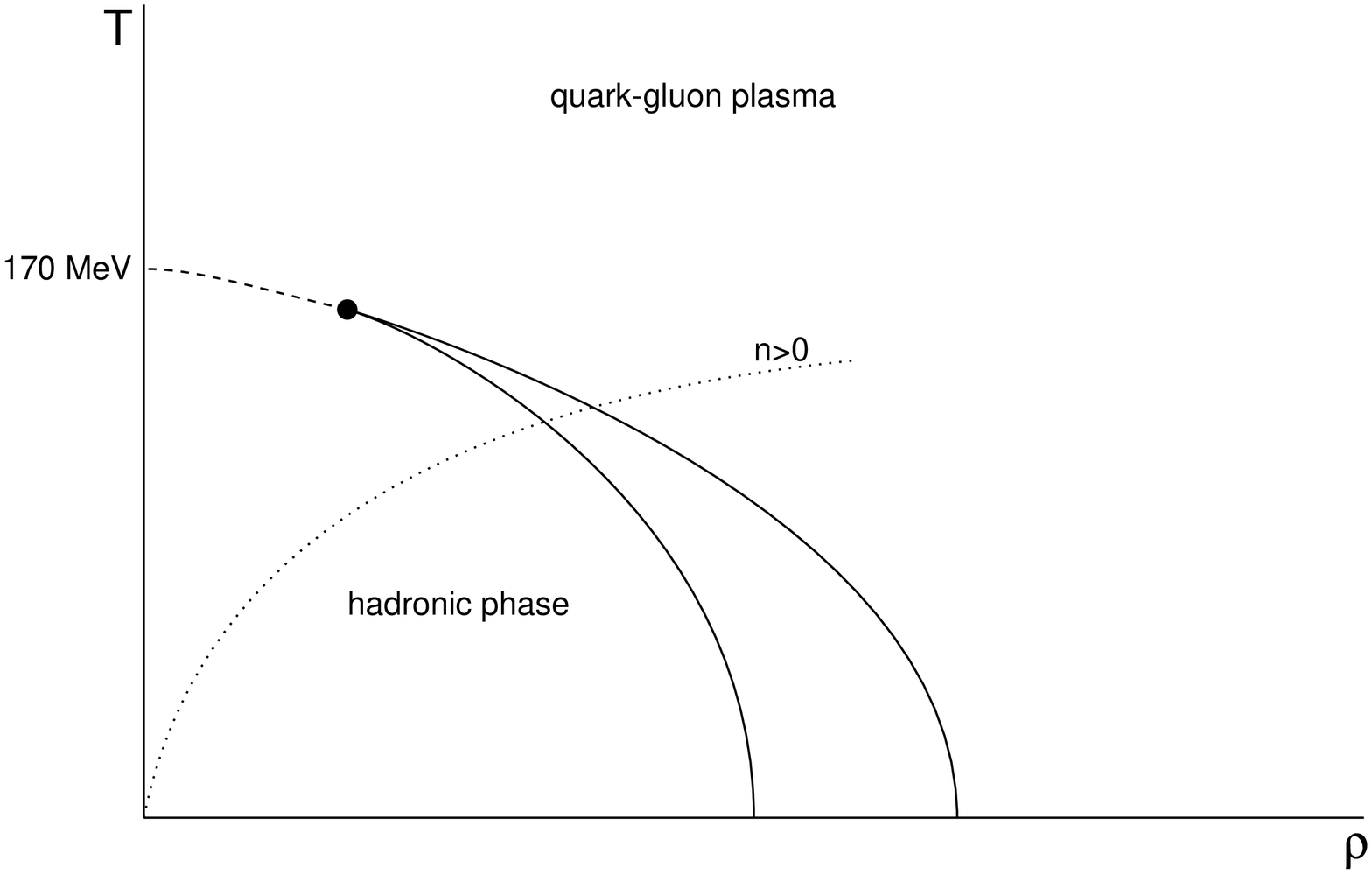}
\caption{A schematic view of the expected QCD phase diagram in the 
temperature-density plane \cite{eng99}. The dotted line, $n>0$, 
represents the trajectory in the
phase space when we keep the baryon number fixed and vary $\beta$.
 \label{fig-schem2}}
\end{figure}

The most straightforward way to look for a deconfining transition is to 
measure the Polyakov loop. Although the average value is expected to vanish
due to the $Z_3$ symmetry, we can look at the average absolute value. This
is expected to increase sharply as we go from the confined to the deconfined
phase. To measure the Polyakov loop we need to fold in the phase
\beq
\left< |P| \right> = \frac{ \left< |P| \alpha \right>_0}
{\left< \alpha \right>_0},
\eeq
where we denoted with $\left<\right>_0$ the average over generated
ensemble.  In Fig. \ref{fig-poly}, we plot the Polyakov loop for our
three sets of simulations as a function of temperature. We see that a
transition occurs somewhere around $170\MeV$ for zero density and, as
we increase the density, the transition becomes less sharp and moves
to lower temperature. This picture agrees with the expectations from a
study with static quarks \cite{eng99}, since at large densities the
transition is expected to be first order and, as a result, the system
will develop a coexistence region. To visualize this, we plot in Fig.
\ref{fig-schem2} the expected phase diagram in the temperature-density
plane.  The main difference from the picture in the temperature -
chemical potential plane (see Fig. \ref{pict-phase-diagram}) is that
the first order transition line is split; we have now a line that
borders the pure hadronic phase and another that borders the pure
quark-gluon plasma phase. In between them, we have a coexistence region
characteristic of a first order phase transition.  As we go
through this region, we expect a more pronounced slope in $|P|$. In the
infinite volume limit, we expect that the slope will change abruptly as
we go through the phase boundaries but we will not see an abrupt
jump in our finite density study.

\subsection{Chemical potential}

In order to compare our results with the results in the grand canonical 
ensemble, we need to measure the chemical potential. The thermodynamic 
definition
\beq
\mu(n) = \frac{\partial F(V,T,n)}{\partial n}=-\frac{1}{\beta}
\frac{\partial\ln \tilde{Z}_C(V,T,n)}{\partial n}
\eeq
would produce
\beqa
\beta\mu(n) &=& i\frac{1}{\tilde{Z}_C(V,T,n)}\int \DU e^{-S_g(U)}  \\
&\times&\frac{1}{N}
\sum_{j=0}^{N-1} \phi_j e^{-i n \phi_j} \det M^2(U_{\phi_j}) =
\left< i\phi \right>_{n}, \nonumber
\eeqa
where $\beta = 1/k_B T$. There are a number of problems with this definition.
Firstly, the partition function is symmetric under the transformation 
$\phi_j \rightarrow \phi_j + 2\pi$. This is not true for this definition of
the chemical potential
and then we can ask why would the chemical potential depend on our choice of
$\phi_j$. Secondly, the chemical potential defined above is the quark chemical
potential; it measures the response of the system when one more quark is 
introduced in the system. If we follow the same logic and measure the 
baryon chemical potential we find that 
$\mu_B(n_B) = \left< i3\phi\right>_{3n_B} = 3\mu(3 n_B)$.
Thus, it seems that the response to introducing a baryon in the system is
linearly related to the quark chemical potential. While this might be true in
the deconfined phase, it is clearly not so in the confined phase. The cost
of introducing one quark in an empty box should be infinite, whereas we expect
that the cost of introducing a baryon should be finite. To address these 
shortcomings, we ``discretize'' the derivative and define the chemical 
potential
\beq
\mu(n) = \frac{F(n+1)-F(n)}{(n+1)-n} = F(n+1)-F(n).
\label{defmu}
\eeq
We see that defined as above, the chemical potential measures the increase 
in the free energy as we add a quark to the system. We find then
\beqa
\mu(n) &=& -\frac{1}{\beta}\ln\frac{\tilde{Z}_C(n+1)}{\tilde{Z}_C(n)} =
-\frac{1}{\beta}\frac{1}{\tilde{Z}_C(n)} \int \DU e^{-S_g(U)} \nonumber \\
&\times&\frac{1}{N}\sum_{j=0}^{N-1} e^{-i\phi_j} e^{-i n \phi_j} 
\det M^2(U_{\phi_j}) \\ 
&=& -\frac{1}{\beta}\left<e^{-i\phi}\right>_n. \nonumber
\eeqa
Similarly, for the baryon chemical potential we find 
\beq
\mu_B(n_B) = -\frac{1}{\beta} \left< e^{-i3\phi}\right>_{3 n_B}.
\eeq
With these new definitions, the quark and baryon chemical potentials are no 
longer linearly 
related and they also satisfy the same symmetries as the partition function.
Moreover, since the partition function for a system with a number of quarks
that is not a multiple of $3$ vanishes when we are in the confined phase, we
have
\beq
\mu(3 n) = -\frac{1}{\beta} \ln \frac{\tilde{Z}_C(3n+1)}{\tilde{Z}_C(3n)}
= +\infty,
\eeq
which is exactly what we expect.

\begin{figure}[t]
\includegraphics[width=6cm, angle=-90]{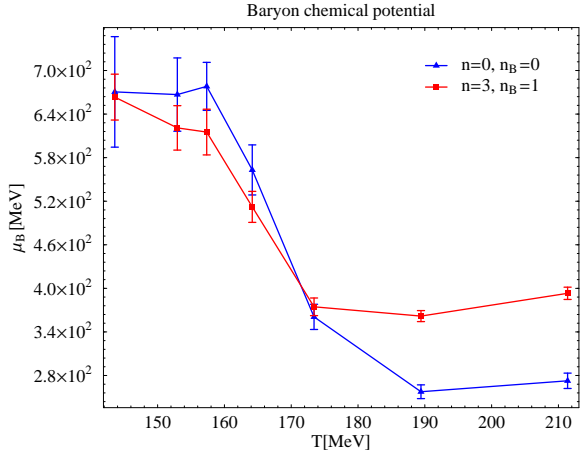}
\caption{Baryon chemical potential as a function of the temperature. We did
not include the data for $n=6$ since it replicates the results for $n=3$.
\label{fig-chem-pot}}
\end{figure}

We note that the chemical potential as defined above has certain symmetries;
since we used $\tilde{Z}_C$ for our definitions we have:
\beqa
\mu(n) &=& \mu(n+N) \nonumber, \\
\mu_B(n_B) &=& \mu(n_B + N/3), 
\eeqa
the second equality holds when $N$ is a multiple of $3$. From charge 
conjugation symmetry, we infer that for $n,n_B>0$
\beqa
\mu(-n) &=& -\mu(n-1) \nonumber \\
\mu_B(-n_B) &=& -\mu_B(n_B-1). 
\label{charge-sym}
\eeqa

In Fig. \ref{fig-chem-pot}, we plot the baryon chemical potential as a
function of temperature. We see that, as we go through the phase
transition, the chemical potential drops sharply. This is due to the
fact that new degrees of freedom become available and the entropy of
the system increases. We notice that in the confined phase, the
chemical potential doesn't change much as we increase the density,
whereas in the deconfined region the chemical potential is larger as
the density increases. These findings are consistent with the results of
Kratochvila and de Forcrand \cite{kra03, kra04}. We would also like to point
out that since in our simulations we used $N=12$, we can show, using the
symmetries of the chemical potential listed above, that $\mu_B(2) = -\mu_B(1)$.
This is why we plot only the curves for $n_B=0$ and $n_B=1$.
				
Computing the chemical potential allows us not only to connect our
results to those from grand canonical simulations, but also to
determine the shape of the phase boundary. To see this, we follow an
argument by Kratochvila and de Forcrand \cite{kra04}. They start by
noticing that the chemical potential in the hadronic phase seems
independent of the baryon number. Based on this observation they build
a simple model where the free energy is proportional to the baryon
number, $F(n_B)=\mu_0 |n_B|$. The coefficient $\mu_0$ is the value of
the chemical potential measured; they show that in this model $\mu_0$
is just the critical chemical potential. Consequently, at any temperature $T$
if we find $\mu$ to be independent of $n_B$ we have determined $\mu_c(T)$. 

\begin{figure}[t]
\includegraphics[width = 6cm, angle = -90]{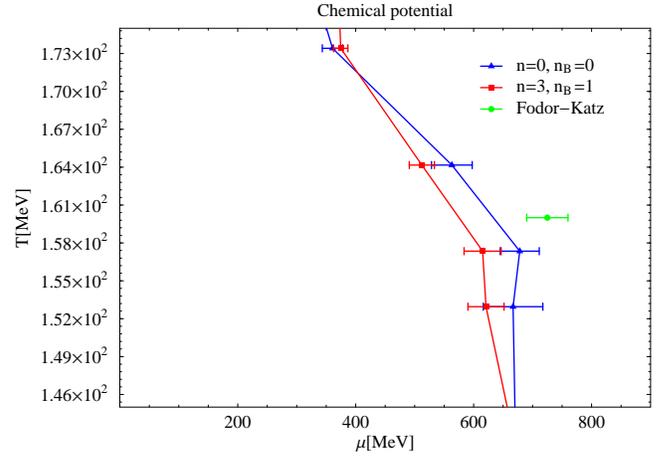}
\caption{Phase transition line based on a model for the free energy.
Note that $\mu$ is the baryon chemical potential.
\label{fig-guess}}
\end{figure}

The argument above can be generalized. In a physical system, the chemical 
potential is
expected to vary from small values at small densities to arbitrarly
large values as the density goes to infinity. However, it can be
argued on general grounds that when the chemical potential stays the
same for a range of baryon numbers we are at a phase transition. In
the thermodynamic limit, the free energy is a convex function of the
baryon number and thus
\beq 
\frac{\partial \mu}{\partial n_B} =
\frac{\partial^2 F}{\partial n_B^2} \geq 0.  
\eeq 
It is then expected that the chemical potential be an increasing
function of baryon number; it will flatten only as we go through the
coexistence region of a first order phase transition. We see then that
we don't really need the free energy to be linear in the baryon
number; when the second derivative vanishes we are at the phase
boundary.

In Fig. \ref{fig-chem-pot} we see that the chemical potential curves for
different baryon numbers overlap, at least at the level of the error
bars, for temperatures lower than $170\MeV$. By the above argument this part
of the curve represents the phase boundary and we can use it to plot 
in Fig. \ref{fig-guess} the phase boundary in the $(T,\mu)$ plane.

Before we move on, we would like to point out that, although not explicit in 
the notation we used above, the chemical potential, as defined in 
Eq. \ref{defmu},
depends also on the volume and the temperature of the system. In the 
termodynamic limit the chemical potential, $\bar{\mu}$, depends only on
temperature and density:
\beq
\bar{\mu}(\rho, T) = \lim_{V\rightarrow\infty}\mu(\rho V, V, T).
\label{muth}
\eeq
It is then more relevant to think of our measured chemical potential as 
being defined at a given density. At small baryon number there is an 
ambiguity as to which density to assign to a particular measurement since
the chemical potential is defined to be the difference of the free energies
at $n_B+1$ and $n_B$ baryon numbers. For large values of $n_B$ this does not
make much of a difference. Most naturally we should think that the measurement
is performed at $n_B+\frac{1}{2}$ and treat $\mu(n_B, V, T)$ as an 
approximation for $\bar{\mu}(\frac{n_B+\frac{1}{2}}{V}, T)$. For example
$\mu(n_B=0, V, T)$ should be thought as approximating 
$\bar{\mu}(\frac{1}{2V},T)$. It is then clear that 
\footnote{We thank F. Karsch for pointing this out and P. de Forcrand for
an interesting discussion on this point.}
$
\lim_{V\rightarrow\infty} \mu(n_B=0, V, T)=\bar{\mu}(0,T)=0,
$
from Eq. \ref{charge-sym} which conforms with expectation.
But this is not the relevant limit; the limit of interest is that of
Eq. \ref{muth}. In other words the density should be kept fixed
when approaching the thermodynamic limit.
Note also that using this convention we can show using 
Eq. \ref{charge-sym} that our approximation for the chemical potential becomes 
symmetric in density, i.e. $\bar{\mu}(-\rho)=-\bar{\mu}(\rho)$.

Another interesting point is that since the chemical potential
$\mu(n_B)$ should decrease as the volume is increased (the density
decreases) it would seem that the phase boundary constructed using the
reasoning we presented above will shift. This is not true: the
argument rests on the fact that the chemical potential stays the same
as we increase the baryon number; we understand that to be a
consequence of the fact that we measure the chemical potential at
densities in the phase coexistence region.  As we increase the volume,
the chemical potential will start to drop only when we get out of the
coexistence region but by then it will no longer be independent of
$n_B$. To get back to a chemical potential that is independent of
$n_B$ we have to increase the baryon number until the density again is
in the coexistence region. Thus the new phase boundary we get at
different volumes should be the same (up to finite volume
corrections).

\subsection{Quark condensate}

As we cross over from the hadronic phase to deconfined phase, we also expect
to restore the chiral symmetry. There is ample empirical evidence that 
the deconfining phase transition and the chiral symmetry restoration occur
at almost the same temperatures. As far as we know there is no theoretical
explanation of this fact, thus it is interesting to see whether this remains
true at finite density. For this purpose, we measure the chiral condensate
$\left<\bar{\psi}\psi\right>$. For fermionic observables, we need not only 
fold in the phase $\alpha(U)$; we also need to perform a separate Fourier
transform. For an arbitrary fermionic bilinear $\bar{\psi}\Gamma\psi$, where
$\Gamma$ is some spinor matrix, we have
\beqa
\left< \bar{\psi}\Gamma\psi\right> &=& \frac{1}{\tilde{Z}_C}\frac{1}{N}
\sum_{j=0}^{N-1} e^{-in\phi_j} \int \DU e^{-S_g(U)} \nonumber \\
&\times&\int \Dpsi e^{-S_f(U_{\phi_j},\bar{\psi},\psi)}
\bar{\psi}\Gamma\psi  \\
&=& \left\langle\sum_{n'=0}^{N-1} \frac{\tdetn{n'}M^2}{\tdetn{n}M^2}
(-2\mbox{Tr}_{n-n'}\Gamma M^{-1})\right\rangle, \nonumber
\eeqa
where the factor of $2$ comes from using two degenerate flavors and we defined
\beq
\mbox{Tr}_n \Gamma M^{-1} \equiv \frac{1}{N}\sum_{j=0}^{N-1} e^{-in\phi_j}
\Tr \Gamma M(U_{\phi_j})^{-1},
\eeq
the $n^{th}$ Fourier component of the trace. Note that when computing a 
fermionic observable, we have contributions not only from the $0^{th}$ 
component $\tdetn{n} M^2 \mbox{Tr}_0\Gamma M^{-1}$, but also from the
parts of the propagator that wrap around the lattice in the time direction.
More importantly, determinant sectors other than $\tdetn{n} M^2$ become 
relevant.

\begin{figure}
\includegraphics[width=6cm, angle=-90]{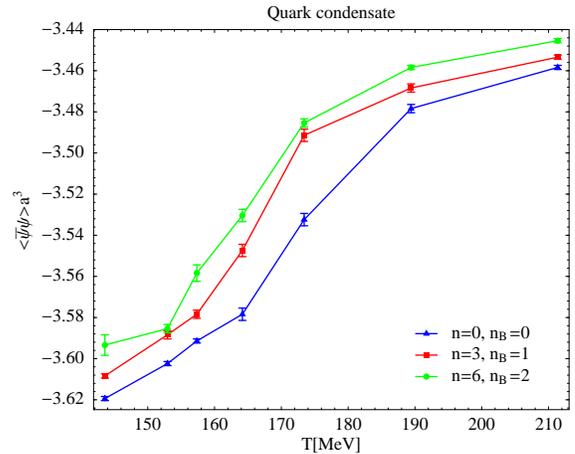}
\caption{The quark condensate $\left<\bar{\psi}\psi\right>$ in lattice units
as a function of temperature. \label{fig-pbp}}
\end{figure}

To look for the chiral restoration phase transition, we measure the chiral
condensate
\beq
\left< \bar{\psi}\psi\right>  \equiv \frac{1}{N_f} \frac{2\kappa}{N_4} \sum_{x}
\left< \bar{\psi}(x)\psi(x)\right>,
\eeq
where $N_f=2$ is the number of flavors and $N_4$ is the lattice four
volume.  In Fig. \ref{fig-pbp}, we plot the quark condensate as
measured in our simulations. We note that as we go through the phase
transition, the quark condensate gets smaller. The slope changes in the
proper temperature range and the transition temperature seem to
decrease with increasing baryon number. Unfortunately, it is
not easy to attach much meaning here, since with Wilson fermions the chiral
symmetry is broken by lattice artifacts and the quark condensate
receives large contributions from these artifacts. Also, the quark mass
we employed is very large so the explicit breaking of the chiral symmetry
is probably large enough to prevent one from seeing any signal in the chiral 
condensate.

\subsection{Conserved charge}

Finally, we turn our attention to the conserved charge. While in the grand
canonical ensemble measuring the conserved charge,
\beqa
Q(t) = -\kappa \sum_{\vec{x}} [\bar{\psi}(x)U_4(x)(1-\gamma_4)\psi(x+\hat{t}) 
\nonumber \\
-\bar{\psi}(x+\hat{t})U_4^\dagger(x)(1+\gamma_4)\psi(x)], 
\eeqa
helps in measuring the average number of particles in the box, there
seems to be little point in measuring the conserved charge in the
canonical ensemble. In fact, we can prove that if you are to use the
true partition function $Z_C$ given in Eq. (\ref{zc}), the charge should
be equal to the number of fermions that we put in the box,
configuration by configuration. However, since we are simulating an
approximation of the partition function, $\tilde{Z}_C$, we can use the
conserved charge to check whether our assumption that $\tilde{Z}_C
\approx Z_C$ is true. It is easy to show that
\beq
\left< Q(t)\right>_{\tilde{Z}_C(n)} = \frac{\sum_m (n+mN) Z_C(n+mN)}
{\sum_m Z_C(n+mN)}.
\eeq
We see then that the deviation from the expected number of quarks would 
quantify how much mixing of different quark sectors we have.

\begin{figure}
\includegraphics[width=6cm, angle=-90]{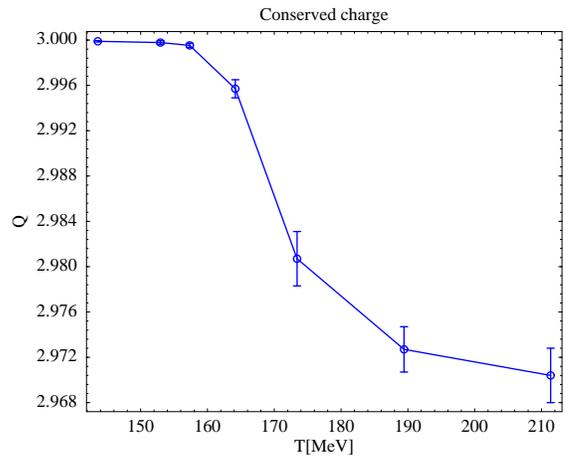}
\caption{Total number of particles in the box for the $n=3$ simulations as
a function of temperature. \label{fig-q}}
\end{figure}

In our simulations, we used $N=12$ and $n=0$, $3$ and $6$. For the $n=0$ and
$n=6$ simulations, we can prove that the conserved charge will be zero.
In the $n=0$ case, this is due to the charge conjugation symmetry 
$Z_C(n) = Z_C(-n)$. For the $n=6$ simulation, this is due to the fact that
for every number of the form $6+12 m$ there is another integer $m'$ such that
$6+12 m' = -(6+12 m)$; plugging this in the expression above we get 
$\left<Q\right>_{n=6} = 0$. The only non-trivial case is when $n=3$, which
we plotted in Fig. \ref{fig-q}. We see that as the quarks become deconfined,
the chemical potential drops and the mixing with the other sectors becomes
more important. However, even for large temperatures, $T\sim 200\MeV$, the
mixing is only about one percent. This implies that even the nearest sector,
$Z_C(n=9)$, is greatly suppressed, i.e. 
$Z_C(n=9)/Z_C(n=3)\sim 0.01$. The mixing is small mainly because the 
chemical potential is large; the mixing will get worse when we 
decrease the fermion mass. 

In conclusion, in this section we show that we can see the expected
deconfining transition in the Polyakov loop, that the chemical
potential drops as we go from the hadronic phase to the quark-gluon
plasma phase, and that there is some hint of the transition even
in the chiral condensate.  Using the conserved charge, we have 
checked that the approximation we made, employing a discrete Fourier
transform rather than a continuous one, is valid.

\section{Conclusions and outlook}

In this paper, we show that the canonical partition function can be used
to investigate the phase structure of QCD at finite temperature and non-zero 
density. The algorithm we employed allows us to investigate densities much
higher that those available by other methods. Sign fluctuations limit our
ability to reach very low temperatures or much larger densities than the
ones explored here. We also checked that the discrete Fourier approximation 
to the canonical partition function introduces only minimal deviations.

The physical picture that emerges from our simulations is consistent
with expectations. The Polyakov loop, the chemical potential and the
quark condensate show signs of a transition around $T\sim
170\MeV$. The quark condensate does not vanish but we need to employ
smaller masses and finer lattices to reasonably expect a clear signal
of chiral symmetry restoration. Another route is to employ a more
sophisticated definition for the chiral condensate, involving perhaps
some form of subtraction, or use chiral fermions.

In the future, we would like to locate points on the phase transition line. 
For this, we need to get to lower temperatures and densities. While we
might be limited in reaching lower temperatures, we should be able to reach
lower densities; all we need is to move to larger volumes. However, since we
have to use larger lattices we need to use an estimator for the determinant.
We should point out that the method used to generate the ensemble has no 
bearing on whether we have a sign problem or not, it is an intrinsic property
of the ensemble. Consequently, the sign oscillations stay the same even when
we employ the determinant estimator. The only thing that is going to change is
the acceptance rate. We anticipate that this should not be a problem since the
acceptance rate is very good for rather large HMC trajectory lengths. However,
this need to be studied further.

Before we conclude, we would like to emphasize that, even if it proves that it
is not feasible to reach lower temperatures, this approach is valuable
since it permits the study at the phase diagram at temperatures close
to $T_c$ and rather large densities. We will then be able to determine
various points on the phase transition line. Much effort is put
nowadays on determining this line, and as we pointed out in the
beginning, the methods used today need to be checked for
reliability. To stress this point, we plot in
Fig. \ref{fig-guess}, next to our phase transition line,
the second order phase transition point as determined
by Fodor and Katz \cite{fod02}. Their simulations use different quark
masses, but the shape of the transition line is expected to change very
little.  Although our error bars are rather large, the plot suggests
the possibility of a discrepancy. This has also been noted in
\cite{kra04} and a possible explanation is provided in
\cite{spl05}. These results seem to indicate a possible overlap
problem. It is then imperative that new simulations are carried out to
check the validity of this important result.  A future study that
employs a determinant estimator will allow us to collect better
statistics and hopefully will settle this question.

\begin{acknowledgements}
The work is partially supported by DOE grants DE-FG05-84ER40154 and 
DE-FG02-95ER40907. The authors thank P. de Forcrand and S. Kratochvila for
useful discussions.
\end{acknowledgements}


\begin{thebibliography}{99}

\bibitem{bar98}
I.M. Barbour, S.E. Morrison, E.G. Klepfish, J.B. Kogut, and M.-P. Lombardo, 
Nucl. Phys. Proc. Suppl. B60 (1998) 220.

\bibitem{cro01}
P.Crompton, Nucl. Phys. B619 (2001) 499.

\bibitem{fod02}
Z. Fodor and S.D. Katz, JHEP 0203 (2002) 014.

\bibitem{fod04}
Z. Fodor and S.D. Katz, JHEP 0404 (2004) 050.

\bibitem{all03}
C.R. Allton, et all, Phys. Rev. D68 (2003)014507.


\bibitem{all05}
C.R. Allton, et all, Phys. Rev. D71 (2005)054508.

\bibitem{for02}
P. de Forcrand and O. Philipsen, Nucl. Phys. B642 (2002) 290.

\bibitem{eng99}
J. Engels, O. Kaczmarek, F. Karsch, and E. Laerman,
Nucl.Phys. B558 (1999) 307-326.

\bibitem{kra03}
S. Kratochvila and P. de Forcrand, hep-lat/0309146.

\bibitem{kra04}
S. Kratochvila and P. de Forcrand, hep-lat/0409072.

\bibitem{liu02}
K.F. Liu, Int. Jour. Mod. Phys. B16 (2002) 2017.

\bibitem{liu03}
K.F. Liu, QCD and Numerical Analysis III, p. 101, Springer, 2005,
hep-lat/0312027.

\bibitem{ale04}
A. Alexandru, M. Faber, I. Horv\'{a}th, K.F. Liu, 
Nucl. Phys. Proc. Suppl. B140 (2005) 517, hep-lat/0410002.

\bibitem{has83}
P. Hasenfratz and F. Karsch, Phys. Lett.B125 (1983) 308.


\bibitem{joo03}
B. Joo, I. Horv\'{a}th, K.F. Liu, Phys. Rev. D67, (2003) 074505.

\bibitem{ale02}
A. Alexandru and A. Hasenfratz, Phys. Rev. D66 (2002) 094502.


\bibitem{fab95}
M. Faber, O. Borisenko, S. Mashkevich, and G. Zinovjev, Nucl. Phys. B42
(1995) 484.

\bibitem{rob86}
A. Roberge and N. Weiss, Nucl. Phys. B275 (1986) 734.


\bibitem{thr98}
C. Thron, S. J. Dong, K.F. Liu, and H.P. Ying, Phys. Rev. D57 (1998) 1642.

\bibitem{som94}
R. Sommer, Nucl. Phys. B411 (1994) 839.

\bibitem{got87}
S. Gottlieb, W. Liu, D. Toussaint, R.L. Renken, and R.L. Sugar, Phys. Rev.
D35 (1987) 2531.

\bibitem{dua87}
S. Duane, A. D. Kennedy, B. J. Pendleton, and D. Roweth, Phys. Lett. B195, 
(1987) 216.

\bibitem{spl05}
K. Splittorff, hep-lat/0505001.


\end{thebibliography}
\end{document}